\documentclass[namedreferences]{SolarPhysics}
\usepackage[optionalrh]{spr-sola-addons}    
\usepackage{url}             

\newcommand{\beq}{\begin{equation}}
\newcommand{\eeq}{\end{equation}}
\newcommand{\beqa}{\begin{eqnarray}}
\newcommand{\eeqa}{\end{eqnarray}}
\newcommand{\beqann}{\begin{eqnarray*}}
\newcommand{\eeqann}{\end{eqnarray*}}

\usepackage{graphicx}
\usepackage{wrapfig}
\usepackage{lscape}
\usepackage{color}

\begin{document}

\begin{article}
\begin{opening}
\title
{On the Asymmetric Longitudinal Oscillations of a Pikelner's Model Prominence}

\author{J.~\surname{Kra\'skiewicz}$^{1}$\sep
        K.~\surname{Murawski}$^{1}$\sep
    A.~\surname{Solov'ev}$^{2,3}$\sep
    A.K.~\surname{Srivastava}$^{4}$\sep
}
\runningauthor{J. Kra\'skiewicz et al.}
\runningtitle{On the Antisymmetric Longitudinal Oscillations of a Pikelner's Model Prominence}


\institute{$^{1}$ Group of Astrophysics, Institute of Physics,
 UMCS, ul. Radziszewskiego 10, 20-031 Lublin, Poland\\
e-mail:kraskiew@umcs.pl
$^2$ Central (Pulkovo) Astronomical Observatory of the Russian Academy of Science, Pulkovskoe Shausse, 65/1, 196140, Saint-Petersburg, Russia\\
$^3$ Kalmyk State University, Elista, 358000, Russia\\
$^4$ Department of Physics, Indian Institute of Technology
(BHU), Varanasi-221005, India\\
}


\begin{abstract}
We present analytical and numerical models of a normal-polarity quiescent prominence that are based on the model
of Pikelner (Solar Phys. 1971, {\bf 17}, 44 ). We derive the general analytical expressions for the two-dimensional equilibrium plasma quantities such as the mass density and
a gas pressure, and we specify magnetic-field components for the prominence, which corresponds to a dense and cold plasma residing in the dip
of curved magnetic-field lines. With the adaptation of these expressions, we solve numerically the 2D, nonlinear, 
ideal MHD equations for a Pikelner's model
of a prominence
that is initially perturbed by reducing the gas pressure at the dip of magnetic-field lines.
Our findings reveal that as a~result of pressure perturbations the prominence plasma starts evolving in time and this leads to
the antisymmetric magnetoacoustic--gravity oscillations as well as to the mass-density growth at the magnetic dip, 
and the magnetic-field lines subside there.
This growth depends on the depth of magnetic dip. For a shallower dip, less plasma is condensed and {\it vice-versa}.
We conjecture that the observed long-period magnetoacoustic--gravity oscillations in various prominence systems are in general
the consequence of the internal pressure perturbations of the plasma residing in equilibrium at the prominence dip.
\end{abstract}
\end{opening}

------------------------------------------------------------------------------------------------
\section{Introduction}

Prominences are dense and cold solar coronal magnetic structures, which
reveal high degree of complexity such as their long and thin threads
(Tandeberg-Hansen, 1974). The average prominence temperature is about two
hundred times lower and the mass density is approximately two hundred times higher than the ambient coronal values.
Prominences are linked to the underlying photosphere by several
foot-points and lie along the polarity inversion line. The length of prominences is within $50-500$ Mm, their height is
$10-100$ Mm, and their average thickness is $15$ Mm (Priest, 1989).

Solar prominences can be classified into two groups: i) active  prominences and
ii) quiescent prominences (Zirin, 1988).  Active prominences reveal
life-times of no more than a few days, undergoing dramatic changes in
plasma motions and magnetic activity. They are often associated with
solar flares. Quiescent prominences can live for
months, forming themselves over a magnetic neutral line that separates 
the regions of opposite magnetic polarities on the photosphere.

The first quiescent prominence models were devised about $60$ years ago (Menzel, 1951).
Two classical models are commonly accepted, which are known as the
Kippenhahn and Schl\"uter (1957), and Kuperus and Raadu (1974) models. In these models the cool and dense plasma is maintained against gravity by the magnetic tension at the local dip of the magnetic arcade. The distribution of
magnetic polarity in this arcade prominence is the same as in the underlying photosphere, {\it i.e.} direct polarity.
Kuperus and Raadu (1974) proposed a model in which the
prominence, taking the form of a magnetic flux rope, is maintained in a vertical current-sheet with open magnetic-field lines. Below this
current-sheet the X-point is present, and the prominence exhibits
inverse polarity. Currently, there are
two well-accepted 3D prominence models with inverse polarity: the sheared arcade and the flux-rope
magnetic structures (Labrosse et al., 2010, and references cited therein).

While these classical models reveal the equilibrium magnetic configurations of quiescent prominences,
a recent trend is emerging with the observational and theoretical reports
that led to the foundation of plasma and wave dynamics within such stable magnetic configurations of the prominence system.
(Luna and Karpen 2012; Luna, D\'iaz and Karpen 2012) have found that the observed large-amplitude longitudinal prominence oscillations ($>20$ km s$^{-1}$)
are driven by the projected gravity along the flux tubes and are strongly influenced by
the curvature of the dips of the magnetic field in which the prominence threads reside.
These oscillations reported by both Luna and Karpen (2012) and  Luna, D\'iaz and Karpen (2012) for slow magnetoacoustic--gravity modes, are
the symmetric longitudinal oscillations.
Terradas et al. (2013) have studied the excitation of fast as well as slow antisymmetric magnetoacoustic--gravity modes
of a prominence; they also dealt with magnetoststatic (MHS) equilibrium and performed linear MHD normal mode analysis.
Small-amplitude longitudinal (slow) and fast transverse prominence oscillations either
as a collective motion or as an individual thread ($\approx 2-3$ km s$^{-1}$) are well observed
and modeled in the solar atmosphere (Arregui, Oliver and Ballester, 2012, and references cited therein).
Terradas, Oliver and Ballester (2001) have reported that slow magnetoacoustic--gravity longitudinal oscillations of
the quiescent prominences can be damped by the radiation and Newtonian cooling.
Longitudinal prominence oscillations were observed and modeled 
by Li and Zhang (2012), Zhang et al. (2012), Shen et al. (2014), Bi et al. (2014), and Chen, Harra and Fang (2014).

While a number of generalizations
to the seminal models of Kippenhan and Schl\"uter (1957) and Kuperus and Raadu
(1974) were constructed in the past, we propose an arcade model of normal polarity as sketched by
Pikelner (1971).
Pikelner (1971) only sketched the model but he did not support it by any analytical expression.
Our  aim is to  devise an analytical arcade model of a prominence
by presenting the stringent mathematical expressions for the equilibrium prominence quantities.
Our devised analytical model of prominence
is significant to the bounded prominence structures
that exhibit collective wave motions as well as the plasma dynamics.
Additionally, our goal is to implement this analytical model into the FLASH numerical code (Lee, 2013)
and develop a numerical model. This numerical model allows us to study a number of plasma phenomena. 
As a particular application of this model, in the present paper, 
we simulate magnetoacoustic--gravity waves in the prominence that result from a localized pressure pulse. 
This is the first article describing in detail our newly developed analytical model 
and its one numerical experiment on long-period prominence oscillations to shed light on their driving physical mechanism. 
The derived results are applicable to understand the physical processes of the prominence during evolved non-linear oscillations and are
thereby useful for prominence seismology and related observations.

This article is organized as follows:
Section~2 describes the analytical model of a
quiescent prominence that is based on Pikelner's model. The results of numerical simulations are outlined in Secion~3.
In the last section we present the discussion and conclusions.

%
\section{Analytical  Model of Quiescent Prominence}\label{sec:ana_model}

\subsection{MHD equations}\label{sec:equ_model}
We consider a gravitationally stratified and magnetically confined plasma,
which is described by ideal two-dimensional (2D) magnetohydrodynamic (MHD) equations:
%
\beqa
\label{eq:MHD_rho}
{{\partial \varrho}\over {\partial t}}+\nabla \cdot 
(\varrho{\textbf {\textit V}})&=&0\, ,\\
\label{eq:MHD_V}
\varrho{{\partial {\textbf {\textit V}}}\over {\partial t}}+ \varrho\left ({\textbf {\textit V}}\cdot \nabla\right )
{\textbf {\textit V}} &=& -\nabla p+ \frac{1}{\mu} (\nabla\times{\textbf {\textit B}})\times{\textbf {\textit B}} +\varrho{\textbf {\textit g}}\, , \\
\label{eq:MHD_B}
{{\partial {\textbf {\textit B}}}\over {\partial t}} &=& \nabla \times ({\textbf {\textit V}}\times {\textbf {\textit B}})\, , \\
\label{eq:MHD_divB}
\nabla\cdot{\textbf {\textit B}} &=& 0\, , \\
\label{eq:MHD_p}
{\partial p\over \partial t} + {\textbf {\textit V}}\cdot\nabla p &=& -\gamma p \nabla \cdot {\textbf {\textit V}}\, ,\\
\label{eq:MHD_CLAP}
p &=& \frac{k_{\rm B}}{m} \varrho T\, .
\eeqa
%
Here ${\varrho}$ is the mass density, $p$ a gas pressure, ${\textbf {\textit V}}=[V_x, V_y, 0]$, ${\textbf {\textit B}}=[B_x, B_y, 0]$, and ${\textbf {\textit g}}
=[0,-g,0]$ represent the plasma velocity, the magnetic field and gravitational
acceleration, respectively.
The value of $g$ is equal to $274$ m s$^{-2}$.
In addition, $T$ is the plasma temperature,  
$\gamma=5/3$ is the adiabatic index, $\mu $ is the magnetic permeability of the plasma, 
and $m$ is a particle mass that is specified by mean molecular weight of $0.6$. Although this value is valid for fully ionized plasma 
of the corona we believe that its larger value will not  qualitatively change the response of the system.
We assumed that $z$ is an invariant coordinate ($\partial/\partial z=0$) and we set the $z$-components of velocity and magnetic field equal to zero. 
This assumption removes Alfv{\'e}n waves from the system in which magnetoacoustic--gravity waves are able to propagate.
\subsection {Equilibrium Configuration}\label{sec:equil}
%
We assume that the solar atmosphere is in static equilibrium ${\textbf {\textit V}}_{e}={\bf 0})$.
It follows from Equations (\ref{eq:MHD_rho})--(\ref{eq:MHD_p}) that in such
model this equilibrium is described by
%
\beqa
\label{eq:st_p}
-\nabla p_{\rm{e}}+\frac{1}{\mu}(\nabla\times{\textbf {\textit B}}_{\rm{e}})\times{\textbf {\textit B}}_{\rm{e}}+\varrho_{\rm{e}}{\textbf {\textit g}}={\textbf {\textit 0}},\\
\label{eq:st_B}
\nabla\cdot{\textbf {\textit B}}_{\rm{e}}= 0.
\eeqa
%
Here the subscript $_{\rm{e}}$ corresponds to the equilibrium quantities.

As a result of Equation~(\ref{eq:st_B}), the magnetic field,
 ${\textbf {\textit B}}_{\rm{e}}$,
can be expressed with the use of the magnetic-flux function,
 $A(x,y)$, as
\beq
{\textbf {\textit B}}_{\rm e}(x,y)=\nabla\times\left(A\hat{\textbf {\textit z}}\right)
\label{eq:Be}
\eeq
%
%
with $\hat{\textbf {\textit z}}$ being a unit vector along the $z$-direction. As a result, the $x$- and $y$-components of the magnetic field are

\beq
\label{eq:BxBy}
B_{\rm{ex}}=\frac{\partial A}{\partial y},\,\hspace{3mm}B_{\rm {ey}}=-\frac{\partial A}{\partial x}.
\eeq

Under the above assumptions, Equations (\ref{eq:MHD_CLAP}) and (\ref{eq:st_p}) are reduced to the following expressions (Low, 1975, 1980):

\beqa
\label{eq:st_px}
\nabla^2 A(x,y) &=& -\mu\frac {\partial p_{\rm{e}}(A,y)} {\partial A},  \\
\label{eq:st_py}
\varrho_{\rm{e}} (x,y) g &=& - \frac {\partial p_{\rm{e}}(A,y)} {\partial y}.
\eeqa
Now, we can specify the RHS of Equation (11) and thereafter
proceed to solve this equation for the magnetic field, however, this is a formidable nonlinear Dirichlet
problem to solve (Low, 1975, 1980). Here we adopt the approach proposed
by (Low, 1980, 1981, 1982) who reversed the mathematical problem by specifying the magnetic
field first and then deriving the equilibrium conditions for the mass density
and the gas pressure. By defining the magnetic flux function, $A(x, y)$, we can
in general integrate Equation (11) to find formula for the equilibrium gas pressure
and then using Equation~(\ref{eq:st_py}) calculate the mass density. Following this
idea, we integrate Equation (\ref{eq:st_px}) regarding $y$-coordinate as a fixed
parameter (Solov'ev, 2010). Then, a variation of $A$ is
\beqa
{\mathrm d}A=\frac {\partial A}{\partial x} {\mathrm d}x+\frac {\partial A}{\partial y} {\mathrm d}y+\frac {\partial A}{\partial z} {\mathrm d}z=\frac {\partial A}{\partial x} dx,
\eeqa
as a result of $\partial A/\partial z=0$ and ${\mathrm d}y=0$.

Using this expression in Equation~(\ref{eq:st_px}) we obtain
\beq
\mu p_{\rm{e}}=-\int\left(\frac {\partial^2 A}{\partial x^2} +\frac {\partial^2 A}{\partial y^2}\right)\frac {\partial A}{\partial x} {\mathrm d}x + \mu C(y),
\eeq
where $C(y)$ is an integration function. As the integration over the first integrand can be performed explicitly, therefore, we obtain
\beq
\label{eq:pe1}
\mu p_{\rm{e}}= - \frac{1}{2}\left(\frac {\partial A}{\partial x}\right)^2-\int \frac {\partial^2 A}{\partial y^2}\frac {\partial A}{\partial x}  {\mathrm d}x+ \mu C(y).
\eeq
%
\begin{figure*}[!ht]
    \begin{center}
    \includegraphics[width=8.0cm, angle=0]{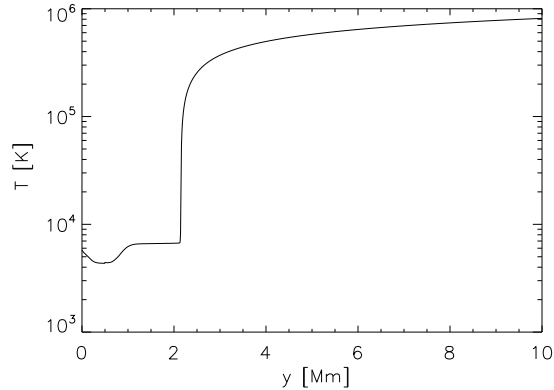}\\
\vspace{-0.5cm}
        \caption{\small Vertical hydrostatic temperature profile from Avrett and Loeser (2008).}
        \label{fig:h_stat_temp}
    \end{center}
\end{figure*}
%
As the magnetic field tends to zero at $x\to \pm \infty$
and $p_{\rm e}=p_{\rm h}(y)$
we find
\beq
C(y)=p_{\rm{h}}(y),
\eeq
where $p_{\rm{h}}(y)$ is the hydrostatic pressure of the magnetic-free solar atmosphere with
\beq
-\frac{\partial p_{\rm{h}}}{\partial y}=\varrho_{\rm{h}} (y) g,
\eeq
which is specified by a temperature profile $T_{\rm e}(y)$.
In this case we adopt the temperature model  of  Avrett and Loeser (2008).
Note that temperature reaches its minimum of 4300 K at $approx 0.6$ Mm (Figure~\ref {fig:h_stat_temp}).
At higher altitudes, the temperature rises with $y$.
At the transition region that is located at $y\approx 2.1$ Mm, the
temperature shows abrupt growth of about 0.8 MK in the solar corona at $y=10$ Mm.
The temperature profile determines uniquely the hydrostatic mass density and a gas pressure,
which fall off with $y$ (not shown here). For a recent implementation of $p_{\rm h}(y)$ see Murawski et al. (2013). From Equation~(\ref{eq:pe1}) we get
\beq
\label{eq:pe}
 p_{\rm{e}}(x,y)=p_{\rm{h}}(y)-\frac{1}{\mu}\left(\int \frac {\partial^2\! A}{\partial y^2}\frac {\partial A}{\partial x} {\mathrm d}x +
\frac{1}{2} \left(\frac {\partial A}{\partial x}\right)^{\!2}\right).
\eeq

From Equation~(\ref {eq:st_py}) it follows that in order to find the distribution of the mass density, $\varrho_{\rm e}(x,y)$, 
we need to calculate $\partial p_{\rm{e}}(A,y)/\partial y$.
In order to execute this action, we perform the following analysis.
Let $A(x,y)$ be a function of independent variables, $(x, y)$.
Then we can express any differentiable function, $S(x,y)$, as $S(y,A(x,y))$
and the following equation is derived:
\beqa
\label{eq:dSdy}
\frac{\partial S(x,y)}{\partial y}=\frac{\partial S(y,A)}{\partial y}+\frac {\partial A}{\partial y}\frac{\partial S(y,A)}{\partial A}.
\eeqa
Then
\beq
\label{eq:dSdy1}
\frac{\partial S(y,A)}{\partial y}=\frac{\partial S(x,y)}{\partial y}-\frac{\partial A}{\partial y}\frac{\partial S(y,A)}{\partial A}.
\eeq
Replacing $S(y,A)$ by $p_{\rm{e}}(y,A)$ and utilizing   Equation~(\ref{eq:st_px}) we find
\beq \label{eq:dSdy2}
\frac{\partial p_{\rm{e}}(y,A)}{\partial y}=\frac{\partial p_{\rm{e}}(x,y)}{\partial y}-
\frac {\partial A}{\partial y} \nabla^2 A,
\eeq
and then expressing $p_{\rm{e}}(y,A)$ by Equation~(\ref {eq:pe})
we find the formula for the equilibrium mass density,
\beq
\label{eq:rho}
\varrho_{\rm{e}} (x,y) g=\varrho_{\rm{h}} (y) g+\frac {1}{\mu}\left[\frac {\partial}{\partial y}\left(\int \frac {\partial^2 \! A}{\partial y^2}\frac {\partial A}{\partial x} {\mathrm d}x +\frac{1}{2}\left(\frac {\partial A}{\partial x}\right)^{\!2}\right)-\frac{\partial A}{\partial y}\nabla^2 A\right].
\eeq

Note that $p_{\rm{e}}$ and $\varrho_{\rm{e}}$ are specified by Equation~(\ref {eq:pe}) and Equation~(\ref {eq:rho}), respectively.
These equations are general in nature.
Specific forms of $p_{\rm e}$ and $\varrho_{\rm e}$ are obtained after the estimation of $A(x,y)$, which is a free function.
This function must be chosen by some physical reasons.
In the following part of this article we present this function for a solar prominence.

\subsection{Pikelner's Prominence Model}
For the solar prominence model of Pikelner (1971) we adopt the following expression for the flux function:

\beq\label{eq:A_Pik}
A(x,y) = \frac {B_0} {k}(1 + a_1 k^2 x^2)\exp[-k^2(a_2 (y-y_{\rm ref})^2+x^2) ]\, ,
\eeq
%
where $a_1$ and $a_2$ are some positive constants, $k$ denotes inverse length, which determines the spatial scale of the structure, and $B_0$ is the magnetic-field strength at the reference point, ($x=0, y=y_{\rm ref}$).
We choose and hold fixed $y_{\rm ref}=10\, {\rm Mm},\, k=1/50\, ({\rm Mm})^{-1}$, and $B_0\approx 6\, {\rm Gauss}$. 
This choice of $A(x,y)$ leads to magnetic-field lines that are characteristic
for a prominence and the equilibrium mass density and gas
pressure can be given by relatively simple expressions.
\begin{figure*}[!ht]
\begin{center}
{\includegraphics[width=8.0cm, angle=0]{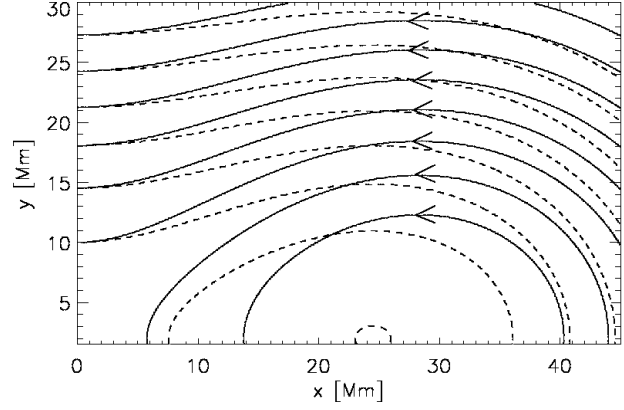}}
\caption{\small Spatial profiles of
magnetic-field lines in Pikelner's prominence model for  $a_2=1.5$, $a_1=1.6$ (dashed lines) and $a_1=2.0$ (solid lines).
Note that only the right-hand ($x\geq 0$) side of the prominence is displayed.}
\label{fig:mag}
\end{center}
\end{figure*}

The magnetic field resulting from Equations~(\ref {eq:A_Pik}) and (\ref{eq:BxBy}) is illustrated in Figure~\ref{fig:mag}. The dip in the magnetic-field vectors is discernible along the vertical line $x=0$, and this dip grows with $a_1$ as magnetic lines are more tilted for a large value of $a_1$ Figure~\ref{fig:mag}).
Having specified $A(x,y)$ by Equation~(\ref{eq:A_Pik}), with the use of Equations~(\ref{eq:pe}) and (\ref{eq:rho}), we express the equilibrium gas pressure and the mass density for the Pikelner's model as

\beqa\label{eq:eq_press}
  \nonumber  \lefteqn{p_{\rm e}(x,y)=p_{\rm h}(y)-}\\
\nonumber & & \mbox{}- 0.5 \, [(p_3 x^4+p_4 x^2+p_7) y^2+p_9 x^4 x^2+p_1 x^4+ p_2 x^2-2 a_2^2]\times\\
& & \times \exp[p_{10} (y-y_{\rm ref})^2+p_{11} x^2] B_0,
\eeqa
\beqa\label{eq:eq_dens}
\nonumber \lefteqn{\varrho_{\rm e}(x,y) =\varrho_{\rm h}(y)+}\\
& &\mbox{}+ 4 \, (p_5 x^4+p_6 x^2+p_8) (y-y_{\rm ref})
\exp[p_{10} (y-y_{\rm ref})^2+p_{11} x^2] B_0/g,
\eeqa
where
\beqann
 p_1  =k^4(-2  a_1^2 a_2-8  a_1^2+8  a_1), & & p_2  =k^2(-4  a_1 a_2+4 a_1^2-8  a_1+4 ), \\
 p_3  =4 k^6 a_1^2 a_2^2,   & & p_4=8 k^4 a_1 a_2^2, \\
 p_5  =k^6(a_1^2 a_2^2-a_1^2 a_2), & & p_6 =k^4[2  a_1 a_2^2+(-a_1^2-2  a_1) a_2], \\
 p_7  =4 k^2 a_2^2, & & p_8 =k^2[a_2^2+(a_1-1 ) a_2], \\
 p_9  =4 k^6 a_1^2, & & p_{10} =-2 k^2 a_2, \; p_{11}  =-2  k^2.\\
\eeqann

\begin{figure*}[!ht]
    \begin{center}
        \includegraphics[width=12.0cm, angle=0]{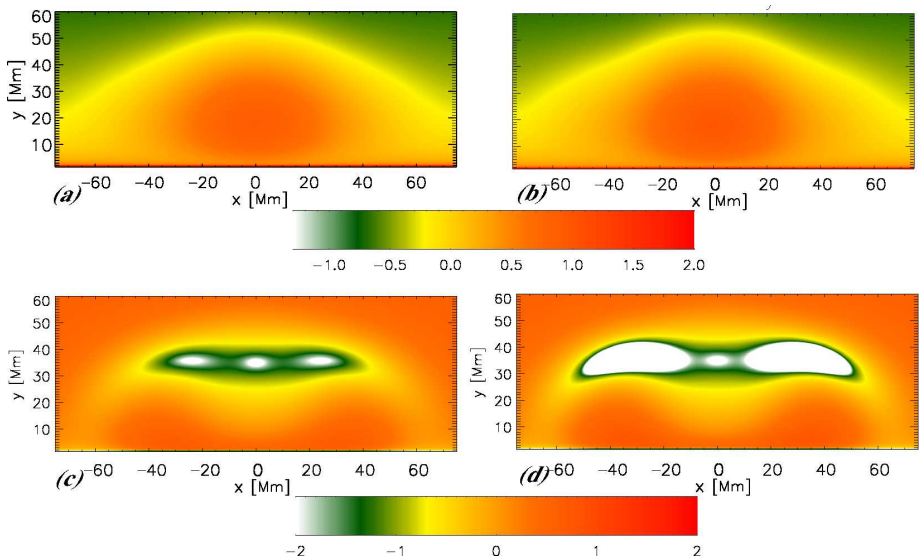}
        \caption{\small Spatial profiles of $\log (\varrho_{\rm e}(x,y))$ for $a_2 = 1.5$, $a_1 = 1.6$ {\bf(a)} and $a_1 = 2.0$ {\bf(b)}. 
                        Temperature profiles $\log (T_{\rm e}(x,y))$ for $a_1 = 1.6$ {\bf(c)} and $a_1 = 2.0$ {\bf(d)}.}
        \label{fig:density}
    \end{center}
\end{figure*}

The prominence equilibrium mass density and temperature profiles that result from Equations~(\ref{eq:eq_press}), (\ref{eq:eq_dens}), and (\ref{eq:MHD_CLAP}) are displayed in Figure~\ref{fig:density}.
Note that the prominence plasma occupies the dense (at $x=0$, $y\approx 20$ Mm) and cold (at $y\approx 35$ Mm) region. While the classical models of a prominence exhibit a single cold region that is centrally located, Pikelner's model reveals the central cold region and two side, cold regions (Figure~\ref{fig:density} c and d). These side regions result from faster fall-off in gas pressure than in mass density.
%

\section{Results of Numerical Simulations}\label{sect:num_res}
\begin{figure*}[!ht]
    \begin{center}
                \vspace{-1.0cm}
        \includegraphics[width=8.0cm, angle=90]{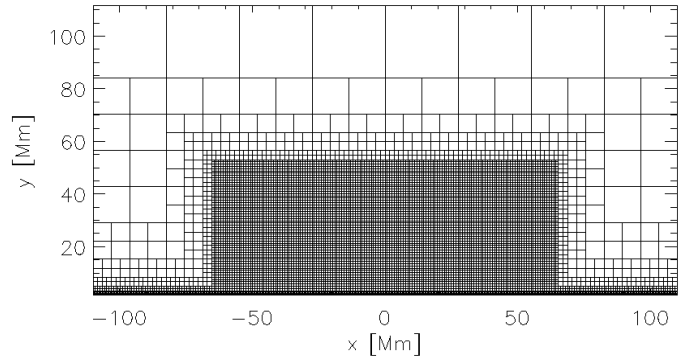}
                \vspace{-2.5cm}
        \caption{\small Block system used in the simulation studies.}
        \label{fig:blocks}
    \end{center}
\end{figure*}
%
Equations (\ref{eq:MHD_rho}) -- (\ref{eq:MHD_CLAP}) are solved numerically using the FLASH code
(Lee and Deane, 2009; Lee, 2013).
This code implements a third-order, unsplit Godunov solver
with various slope
limiters and Riemann solvers ({\it e.g.} T\'{o}th, 2000), as well as adaptive mesh refinement (MacNeice et al., 1999).
We use the minmod slope limiter and the Roe Riemann solver.
For all cases considered  we set the simulation box as
$(-110,110)\, {\rm Mm} \times (1,111)\, {\rm Mm}$ and 
impose boundary conditions by fixing in time all plasma quantities
at all four boundaries to their equilibrium values.
In all of our studies we use a static grid
 with a minimum (maximum) level of refinement set to 3 (9).  The whole computational region is covered  by a set of blocks with different grid cell sizes, which are organized in a hierarchical fashion using a tree data structure Figure~(\ref{fig:blocks}). The blocks at the first/top level of the tree consist of the largest cells, while their children have a factor of two smaller cells and are called to be refined.
As a result, at level $3$ ($9$) the block size contains cells by the factor of $2^2$ ($2^8$) smaller than the top level cells.
In this way, we attain
the effective finest spatial resolution of about $20$ km below the transition region, $y=2.1$ Mm.  The total number of grid cells implemented at $t=0$ seconds in the model is almost $10^6$. The typical
computation time is 48 hours and the computations were performed 
with the $32$ CPUs. 
%
%
\subsection{Perturbations in the Pikelner's Prominence}
We initially perturb the above described equilibrium impulsively by adding a Gaussian pulse in a gas pressure, {\it viz}
\beq\label{eq:perturb}
p(x,y,t=0) = p_{\rm e}(x,y)\left[1+A_{\rm p} \exp\left( -\frac{x^2+(y-y_{\rm 0})^2}{w^2} \right)\right]\, .
\eeq
Here the symbol $A_{\rm p}$ denotes the amplitude of the initial pulse,
$y_{\rm 0}$
its initial position, and $w$ is its width. 
The relative amplitude of the initial pressure pulse, {\it i.e.}, $A_{\rm p} = -0.95$ means that we consider large-amplitude oscillations. 
We set and hold fixed $A_{\rm p}=-0.95\,$, $w=4$ Mm, and $y_{\rm 0}=35$ Mm,  
where $y_{\rm 0}$ corresponds to the position of bottom of the magnetic dip. 
The negative value of $A_{\rm p}$ mimics rapid cooling of plasma, 
which was recently studied by Murawski, Zaqarashvili and Nakariakov (2011).
The sudden, spatially
localized decrease in the gas pressure around the null point 
sucks the plasma in, creating a region of the density enhancement around the $X$-point.
The compression of mass density is subject to buoyancy force, and its influence is expressed by changing the blob shape.

\subsection{Dynamics of the Perturbed Plasma}
\begin{figure*}[!ht]
    \begin{center}
        \includegraphics[width=12.0cm, angle=0]{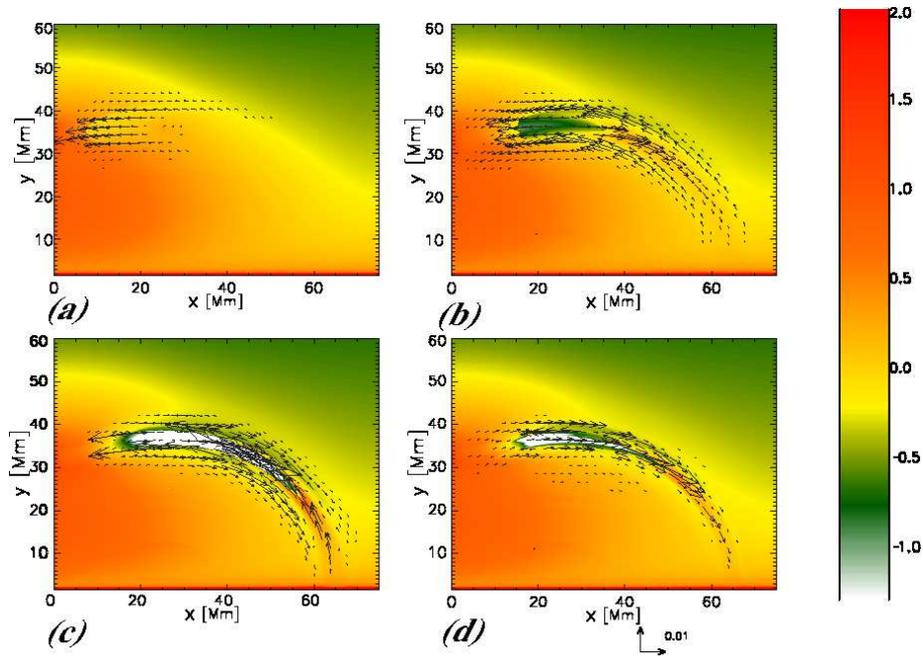}
        \caption{\small Spatial profiles of $\log (\varrho(x,y))$ for $a_1 = 1.6$ and $a_2 = 1.5$ at $t=1000$ seconds {\bf (a)}, 
          $t=2500$ seconds {\bf (b)}, $t=4000$ seconds {\bf (c)}, and $t=5500$ seconds {\bf (d)}.
The vectors denote the velocity of the perturbed plasma. The only $x\geq 0$ Mm is displayed.}
        \label{fig:dens-time-6}
    \end{center}
\end{figure*}
%

\begin{figure*}[!ht]
    \begin{center}
        \includegraphics[width=10.0cm, angle=0]{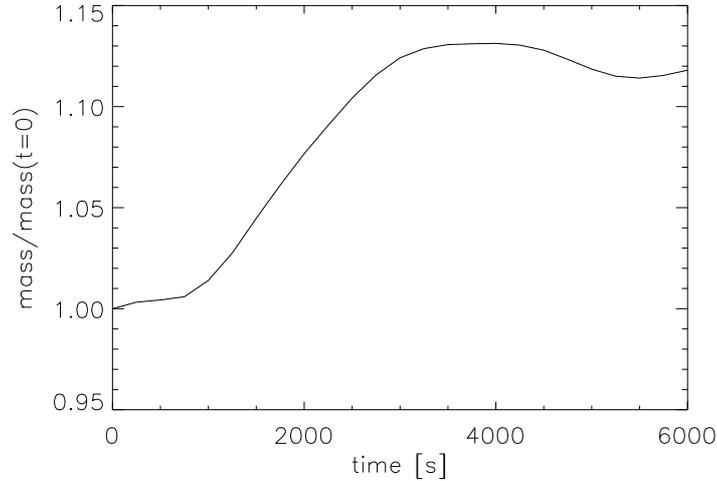}
        \caption{\small The relative mass of the prominence {\it vs}. time}
        \label{fig:mass}
    \end{center}
\end{figure*}
%

\begin{figure*}[!ht]
    \begin{center}
        \includegraphics[width=12.0cm, angle=0]{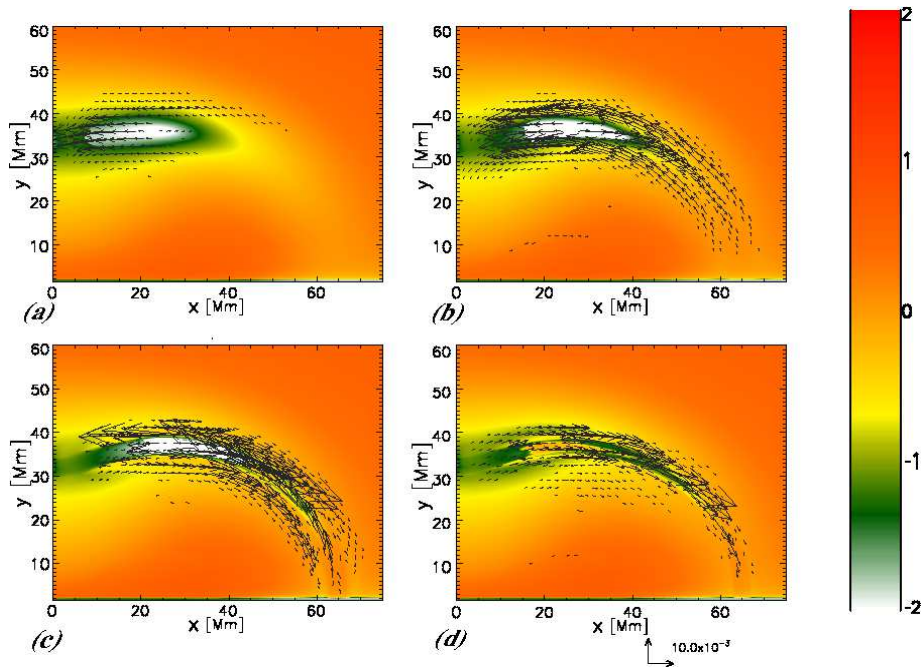}
        \caption{\small Spatial profiles of $\log (T(x,y))$ for $a_1 = 1.6$ and $a_2 = 1.5$ at $t=1000$ seconds {\bf (a)}, 
                        $t=2500$ seconds {\bf (b)}, $t=4000$ seconds {\bf (c)}, and $t=5500$ seconds {\bf (d)}. The velocity vectors are denoted by arrows.}
        \label{fig:temp-time-6}
    \end{center}
\end{figure*}

\begin{figure*}[!ht]
    \begin{center}
     \includegraphics[width=10.0cm,angle=0]{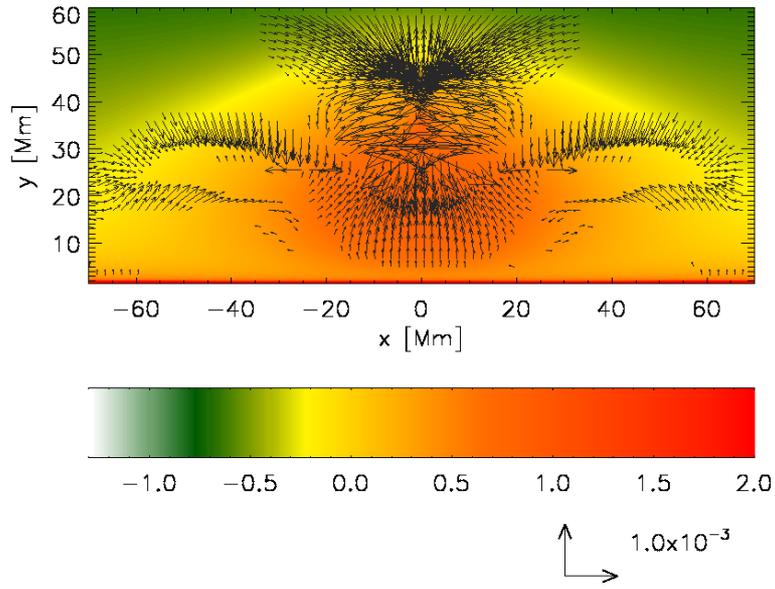}
        \caption{\small {\bf Spatial profiles of $\log (\varrho(x,y))$ for $a_1 = 1.6$ at $t=250$ seconds. The vectors denote the velocity of the perturbed plasma.} }
        \label{fig:rhoVxVy}
    \end{center}
\end{figure*}

\begin{figure*}[!ht]
    \begin{center}
        \includegraphics[width=10.00cm, angle=0]{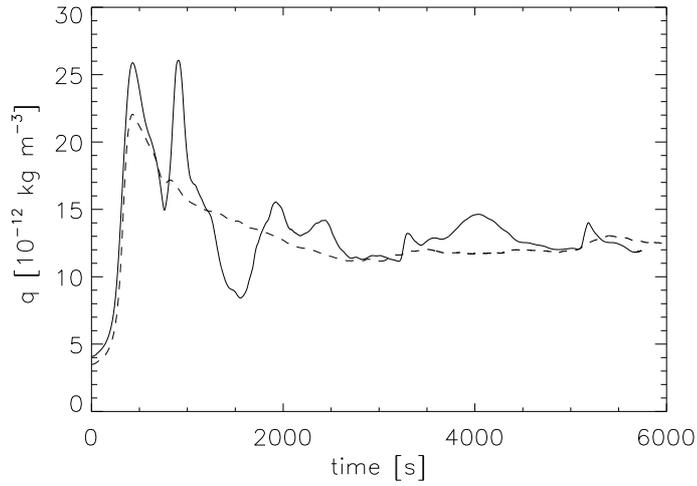} 
        \caption{\small Time-signatures of $\varrho(x=0\,  {\rm Mm},y=35\, {\rm Mm})$ {\it vs.} time for $a_1 = 1.6$ (dashed line)
                         and $a_1 = 2.0$ (solid line).}
        \label{fig:dens-time-a1}
    \end{center}
\end{figure*}

\begin{figure*}[!ht]
    \begin{center}
                \mbox{
        \includegraphics[width=5.75cm,angle=0]{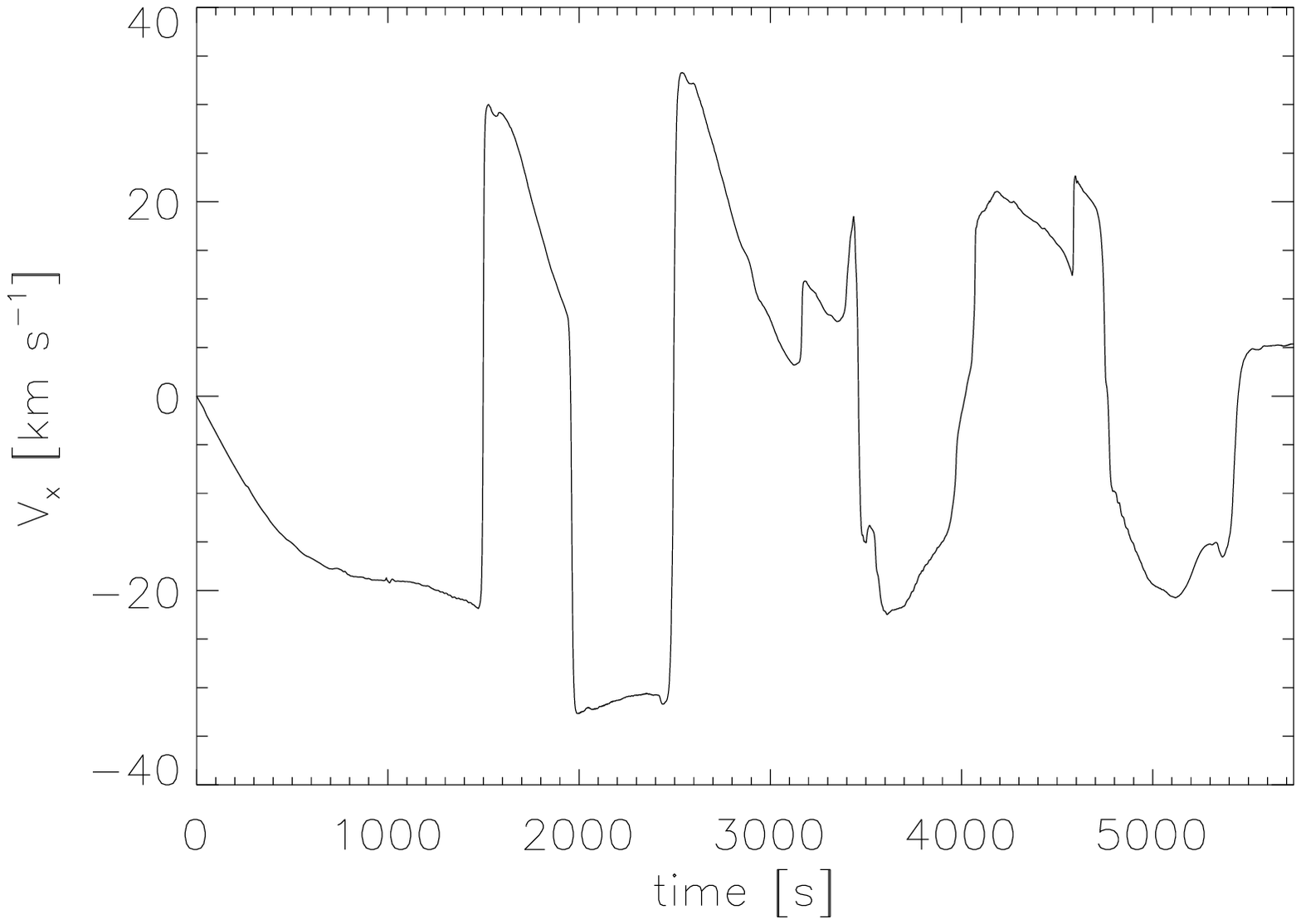} 
        \includegraphics[width=5.75cm,angle=0]{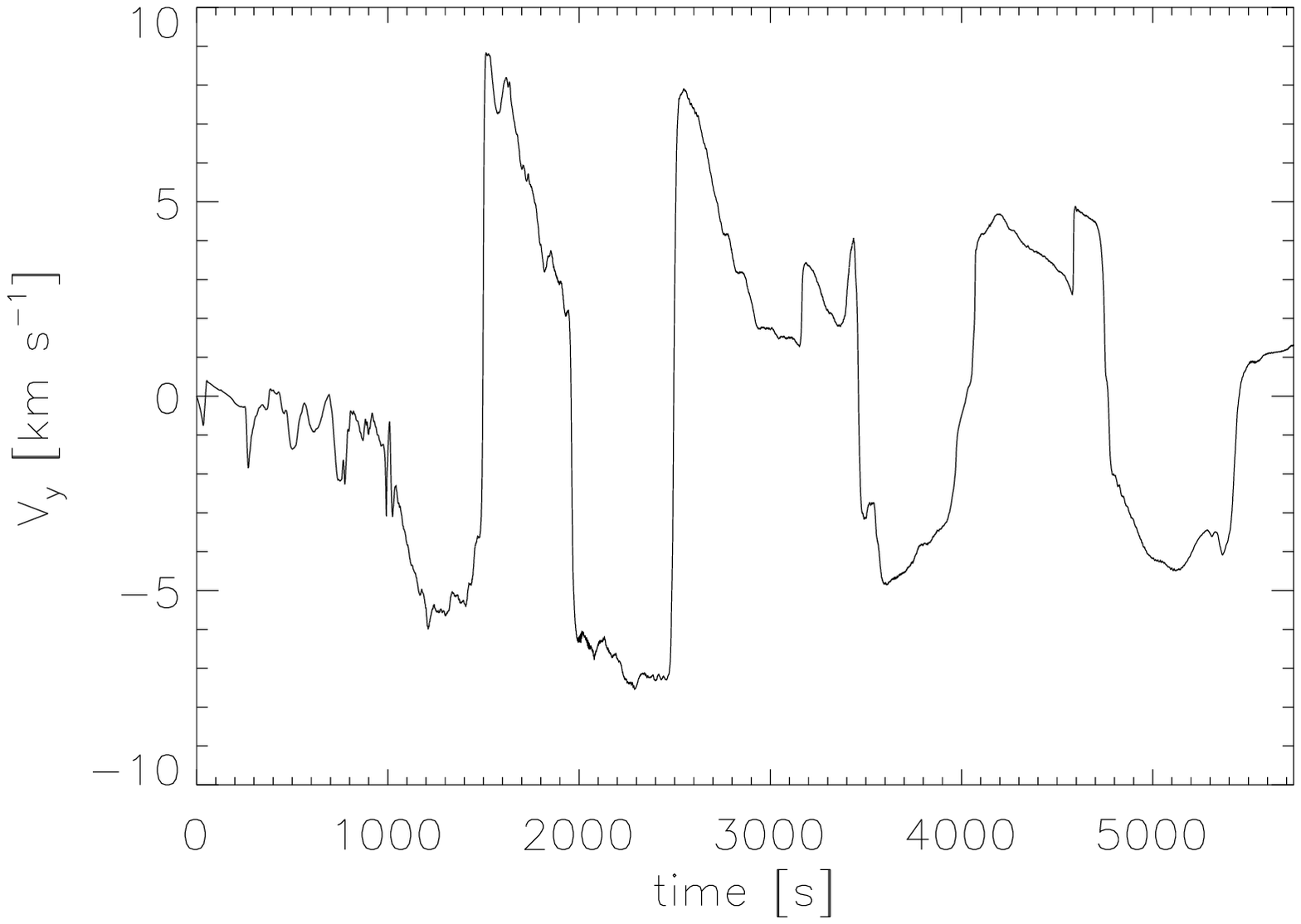}}
        \caption{\small Time-signatures of velocity $V_x(x=20\,  {\rm Mm},y=35\, {\rm Mm})\, {\rm km\, s^{-1}}$ (the left panel) and  $V_y(x=20\,  {\rm Mm},y=35\, {\rm Mm})\, {\rm km\, s^{-1}}$ (the right panel) vs time for $a_1 = 2.0$.
}
        \label{fig:VxVy}
    \end{center}
\end{figure*}
%
\begin{figure*}[!ht]
    \begin{center}
                \mbox{
        \includegraphics[width=5.75cm,angle=0]{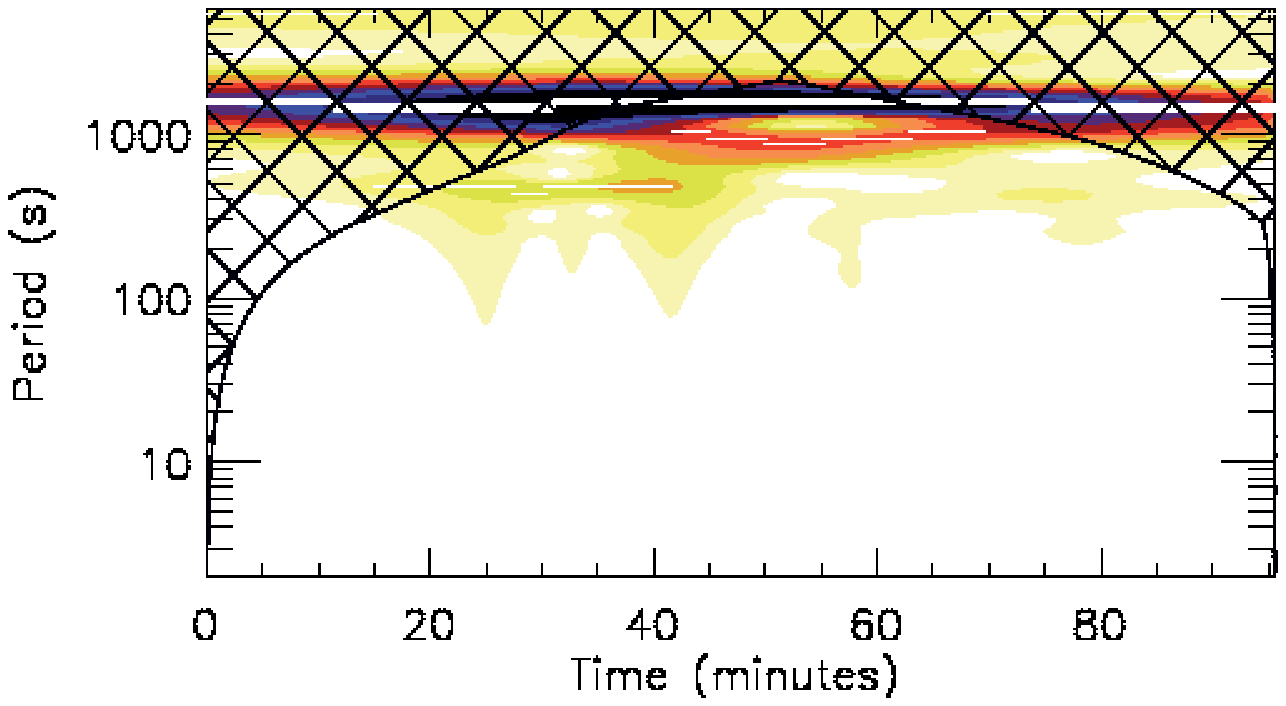} 
        \includegraphics[width=5.75cm,angle=0]{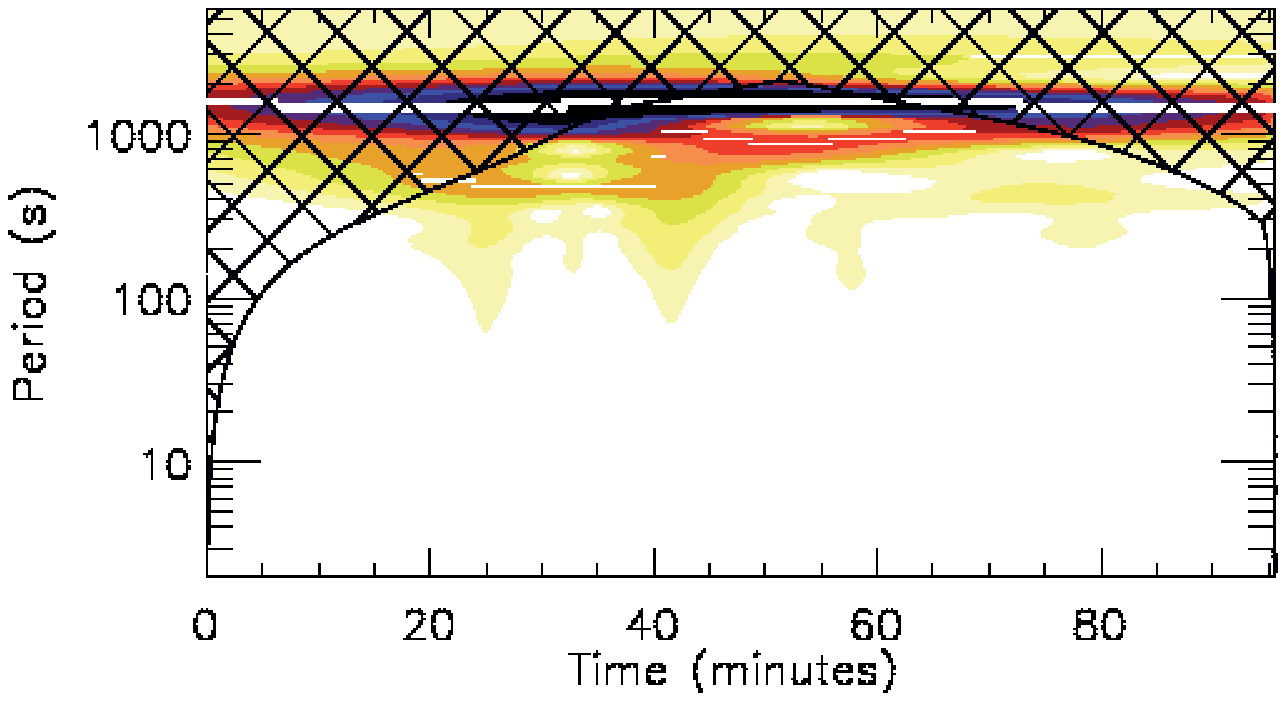}}
        \caption{\small The wavelet spectrum of velocity $V_x(x=20\,  {\rm Mm},y=35\, {\rm Mm})\, {\rm km\, s^{-1}}$ (the left panel) 
          and $V_y(x=20\,  {\rm Mm},y=35\, {\rm Mm})\, {\rm km\, s^{-1}}$ (the right panel) for $a_1 = 2.0$.
}
        \label{fig:WLT_VxVy}
    \end{center}
\end{figure*}
%

\begin{figure*}[!ht]
    \begin{center}
                \mbox{
		  \includegraphics[width=5.75cm,angle=0]{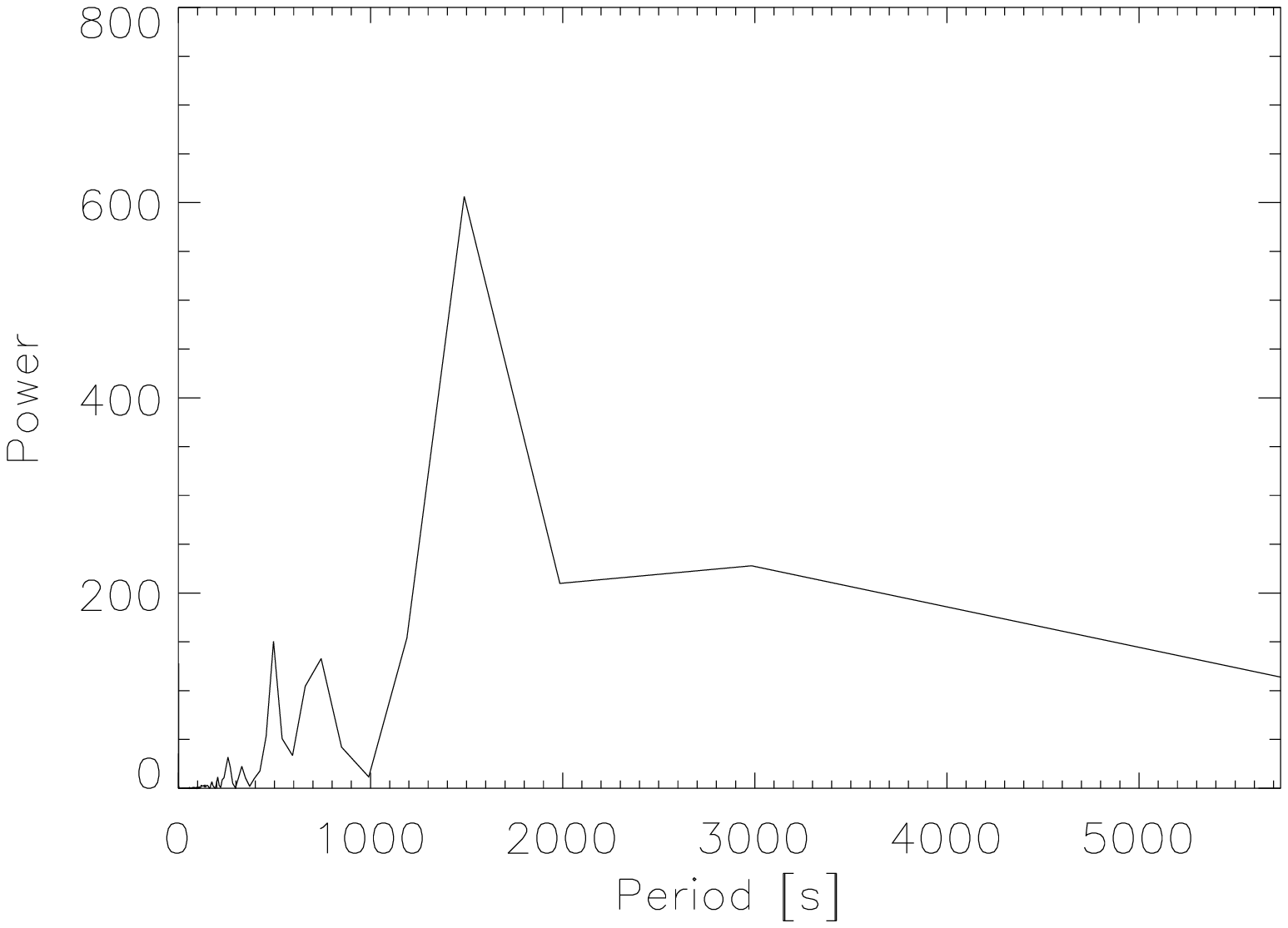}
        \includegraphics[width=5.75cm,angle=0]{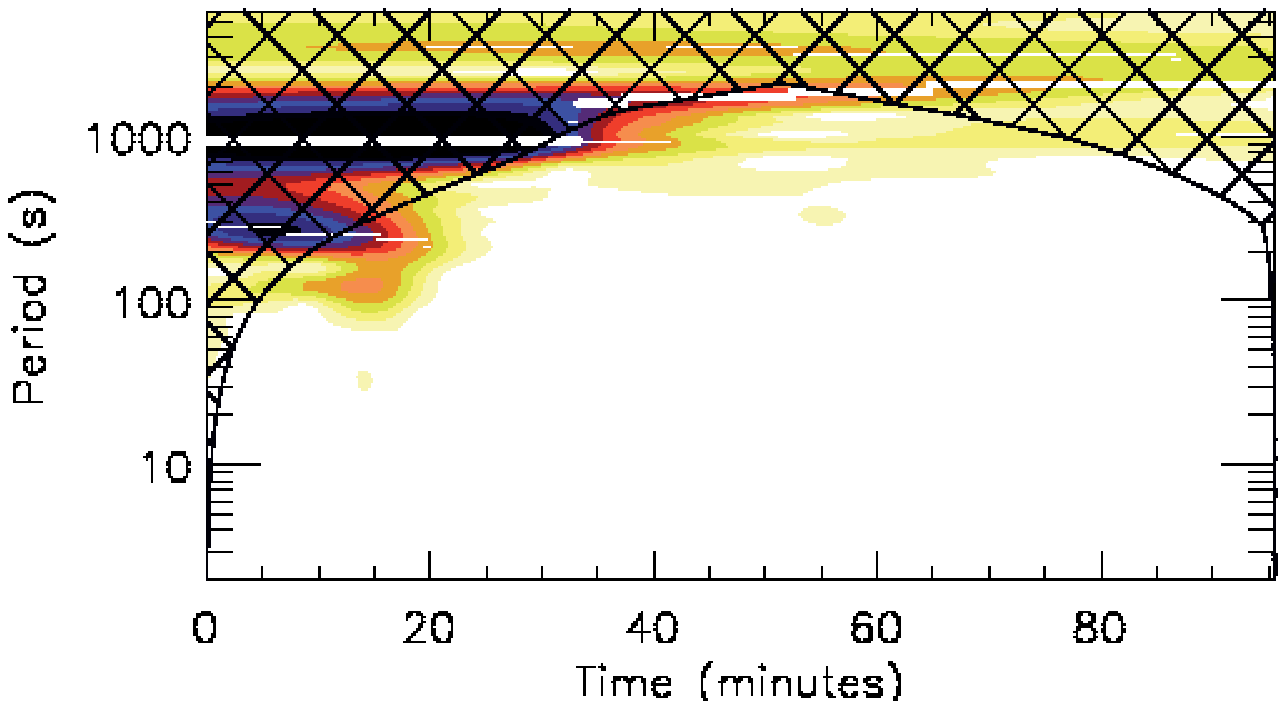} 
        }
        \caption{\small The Fast Fourier Transform (FFT) spectrum of time-signatures of mass density (the left panel) and the wavelet analysis of the density (the right panel) for $a_1 = 2.0$ .
}
        \label{fig:period_rho}
    \end{center}
\end{figure*}

\begin{figure*}[!ht]
    \begin{center}
        \includegraphics[width=8.0cm, angle=0]{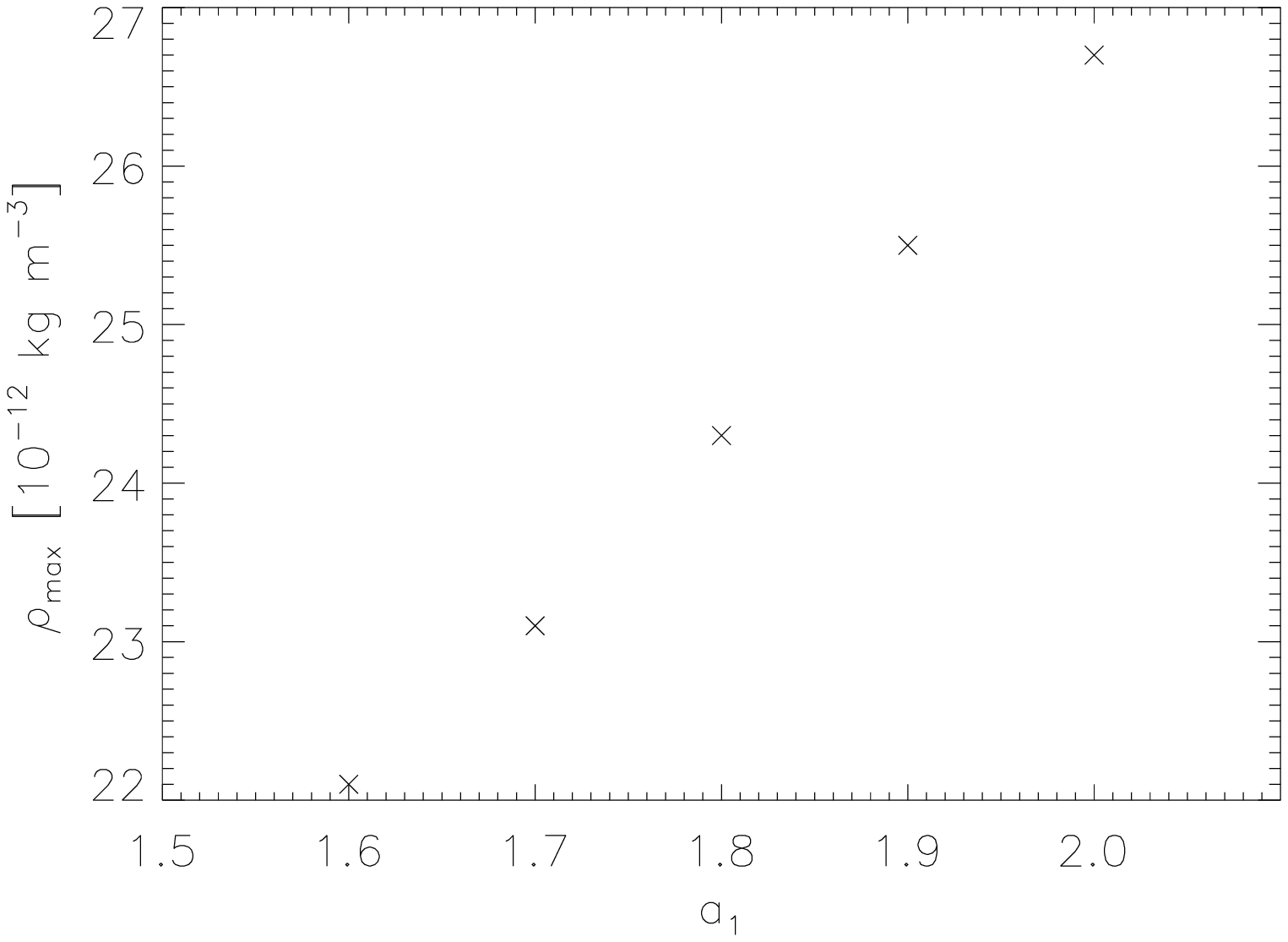}
        \caption{\small Maximum of the mass density taken from the left panel of Figure~\ref{fig:dens-time-a1}
                        {\it vs.} $a_{\rm 1}$.}
        \label{fig:rhomax}
    \end{center}
\end{figure*}

%
Figure~\ref{fig:dens-time-6} displays $\log(\varrho_{\rm e}(x,y,t))$ at four consecutive moments of time.
As a result of the initial perturbation
 magnetoacoustic--gravity waves,
which propagate in the system are essentially excited
within the Pikelner's prominence.
However, these waves quickly leave the prominence domain.
Later the under-pressure that settles at the launching place results in slow magnetoacoustic--gravity waves, which propagate essentially along
the magnetic-field lines (Figure~\ref{fig:dens-time-6} b).
This under-pressure region brought some extra plasma from the ambient region into the launching place (Figure~\ref{fig:dens-time-6} c and d).

Figure~\ref{fig:mass} illustrates the relative mass, $M_{\rm r}$, of the prominence. This mass is evaluated within the region $|x| \leq 20$ Mm, 20 Mm $\leq y \leq$ 45 Mm. Indeed, according to our expectations, this mass grows in time; at $t = 0$ s, $M_{\rm r}=1$ while at $t=4000$ s,  $M_{\rm r}$ reaches a magnitude of 1.13.

Figure~\ref{fig:temp-time-6} shows temperature profiles at four
instants of time.
We see that plasma at the top of the Pikelner's arcade becomes colder than in the ambient medium.
As we are considering
adiabatic equations, the cooling plasma results
from the inflow of plasma,
which is caused by the initial depression in the gas pressure there.
As a result of this depression plasma is attracted by the pressure gradient force into the launching place.

Note that we do not include any dissipation mechanism in our model, such as radiation or thermal conduction.
The initial negative pressure imbalance produces an increase of the mass density at the dip of the model
prominence because it produces an inflow there 
mainly from chromospheric plasma, which is seen in Figure~\ref{fig:rhoVxVy}
after the first 250 seconds of the simulation. 
As the dip is a potential well,
some part of the inflowing plasma is trapped there and the mass density grows there.
A similar physical
explanation can be applied to the temperature field of the prominence.
The plasma flows from the foot-points of the prominence along its field lines,
which is colder than the plasma in the dip at $t=0$ seconds.

Figure~\ref{fig:dens-time-a1} displays the temporal signatures that are drawn by collecting
wave signals in the mass density at the point $(x=0, y=35)$ Mm,
which corresponds to the location of the prominence dip.
These time-signatures reveal an initial phase with strong mass-density variations. These initial phase lasts until $t \approx 1500$ seconds. Later on, the
oscillations are of lower amplitude indicating a strong attenuation. 
The attenuation may
result from energy leakage through the foot-points of the prominence that are located at the
transition region, and also from the dip in magnetic-field lines. We will discuss some estimations to support
the inherent wave-leakage process to dissipate the magnetoacoustic oscillations in the later part of the manuscript.
The temporal evolution of the mass density exhibits a clear non-linear
behavior because the initial mass density, {\it i.e.} the equilibrium value, is smaller than
the ending value, which remains almost constant. The velocity field also oscillates
according to the mass-density perturbations. 

Figure~\ref{fig:VxVy} shows the velocity oscillation 
near the maximum of the prominence at ($x=20,\, y=35$) Mm. 
The velocity-field perturbations are aligned with the magnetic-field lines indicating that the motions are associated with slow modes.

These oscillations for the case of $a_1=2.0$, which represent the slow 
magnetoacoustic--gravity oscillations along the arcade, 
are analyzed by the Fast Fourier Transform (FFT) method. 
They exhibit the main period of $1577$ seconds as is seen in Figure~\ref{fig:WLT_VxVy} and  \ref{fig:period_rho}.

Figure~\ref{fig:rhomax} shows that the maximum of the mass density, $\varrho_{\rm max}$,
grows with $a_{\rm 1}$.
Note that a larger value
of $a_{\rm 1}$ corresponds to
a larger depth of the dips
in the magnetic-field lines;
the prominence with a larger depth of the dips,
once perturbed and departed from the equilibrium,
exhibits a larger amplitude of the slow magnetoacoustic--gravity oscillations.
As a result, more plasma can be trapped in the dips and the mass density becomes larger there,
which explains the growing trend exhibited in Figure~\ref{fig:rhomax}.

%
\begin{figure*}[!ht]
    \begin{center}
        \includegraphics[width=8.0cm, angle=0]{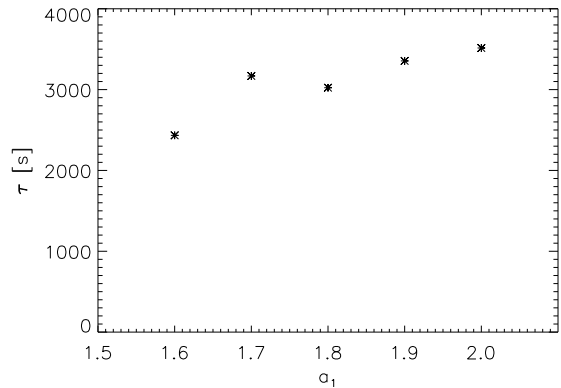}
        \caption{\small Variation of {\bf attenuation time $\tau$ {\it vs.}} $a_{1}$.}
        \label{fig:tau}
    \end{center}
\end{figure*}

From Figure~\ref{fig:dens-time-a1} we infer that
a larger depth of the dips in the magnetic-field lines of
a prominence leads to more oscillations in the mass density.
Compare the solid line drawn for the larger depth in the dip with
the dashed line, which corresponds to the case of the shallower dip.

The decay of the slow magnetoacoustic--gravity oscillations clearly decreases
with
the depth of the magnetic dip of a prominence.
Figure~\ref{fig:dens-time-a1} shows the
fast
attenuation and almost no
oscillatory motion
in the case of a shallower depth of the magnetic dip (dashed line).
Figure~\ref{fig:tau} illustrates how attenuation time, $\tau$, grows with $a_1$. This means 
that a prominence with a larger dip (larger $a_1$) has a longer attenuation time ({\it i.e.}, 
longer decay) and {\it vice-versa}.

%
Terradas et al. (2013)
also found the larger oscillation wave periods for heavy prominences, while lower periods waves
for light prominences. 

\begin{figure*}[!ht]
    \begin{center}
        \includegraphics[width=8.0cm, angle=0]{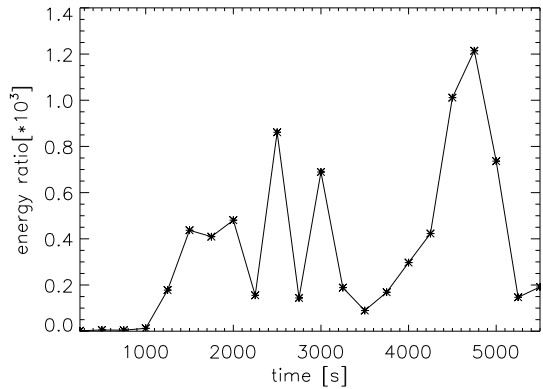}
        \caption{\small The ratio of energy under the foot-points to total energy.}
        \label{fig:energy}
    \end{center}
\end{figure*}
\begin{figure*}[!ht]
    \begin{center}
        \includegraphics[width=8.0cm, angle=0]{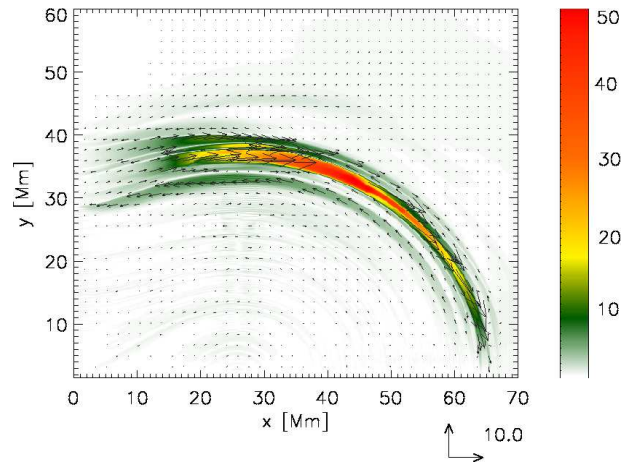}
        \caption{\small Total velocity in km s$^{-1}$ (colour maps and vectors) for $a_1=1.6$ at time $t=6000$ seconds.}
        \label{fig:tot_vel}
    \end{center}
\end{figure*}

Our model deals with the prominence and its ambient medium within the framework 
of the analytical technique of Solov'ev (2010) in which all forces acting on the plasma are taken 
into account. The oscillations excited in the system are not pure gravity or gas pressure, 
which is a different aspect compared to the previous model.
The results in our article are associated with the antisymmetric longitudinal oscillations, which is a case study of our
developed prominence model, and they match the results obtained by Terradas et al. (2013). However, it should be noted that
the present model is based on nonlinear full MHD, while the model by Terradas et al. (2013) deals 
with magnetoststatic equilibrium and linear MHD normal-mode analysis. Therefore, in the context of the 
model description, these two cases have weak relevance.
As stated above, the most likely dissipation mechanism of the excited oscillations is a wave leakage. 
This can result from energy leakage through the foot-points of the prominence, 
which are located at the transition region and from the energy leakage through the 
dip in magnetic-field lines also. Figure~\ref{fig:energy} shows the ratio of kinetic energy of the plasma 
just under the foot-points to the total kinetic energy above the foot-points, {\it i.e.}, above the transition region. 
We see that this ratio grows with time, which indicates the leakage of energy from the prominence. 
Also in Figure~\ref{fig:tot_vel}, we see some leakage of energy under and above the prominence in the form of side streams of fast-moving plasma. 
Therefore, we quantify the presence of wave-leakage process as the dissipative agent for the 
evolved antisymmetric magnetoacoustic--gravity oscillations in Pikelner's model prominence.

We infer from Figure~\ref{fig:tot_vel} that maximal pulse velocity is about $50$ km~s$^{-1}$. 
This magnitude of the flow matches the typical observational data. 
For a large amplitude of the initial pulse, nonlinear effects would become more important. 
However, we have verified by running the appropriate cases that the general scenario of the system evolution remains similar.

%

%


%
\section{Summary}
In this article we adopted the analytical methods of Solov'ev (2010) to derive  Pikelner's model (Pikelner, 1971) of the normal polarity
solar prominence. We implemented our analytical model into the publicly available FLASH code (Lee and Deane, 2009)
and demonstrated the feasibility of fluid simulations in obtaining
quantitative features in weakly magnetized and gravitationally stratified
prominence plasma. In particular, we focused on perturbation of the
developed normal polarity prominence model
and its subsequent evolution. This perturbation was materialized
by launching the initial pulse in gas pressure, which excited
fast and slow magnetoacoustic--gravity waves. Fast waves were present at
the initial stage of the prominence evolution while slow waves
resulted later on and they lead to accumulation of plasma at
the launching place which correspond to location of the magnetic dip.
The parametric studies that we performed revealed that this accumulation
varies with the shallowness of the dip; for a shallower dip there is less
accumulated plasma and smaller oscillations at the magnetic dip.

Such large-amplitude oscillations consist of the motions with the observed velocities 
greater than $20$ km~s$^{-1}$ (Arregui, Oliver and Ballester, 2012), which is also evident in our 
model results. Large-amplitude longitudinal
oscillations can be excited by impulsive events, {\it e.g.}, microflares due to impulsive heating
({\it e.g.}, Zhang et al., 2012).
Therefore in our model, we consider the gas pressure perturbation as a physical initial 
trigger mechanism of the oscillations that excite longitudinal fast and slow magnetoacoustic--gravity waves.

Our new analytical prominence model with realistic
temperature distribution shows reasonable physical
behavior of the typical slow acoustic oscillations in such quiescent prominences, which also matches
the numerical results of Terredas et al. (2013). However, it should be noted that Terradas et al. (2013) deal with magnetoststatic (MHS) equilibrium and linear MHD normal mode analysis, while our model is based on non-linear full MHD equations.
The general damping mechanism may most likely be the radiative cooling
as invoked by many analytical and numerical investigations ({\it e.g.}, Terradas, Oliver and Ballester, 2001; Terradas et al., 2013)
related to the slow acoustic oscillations of the quiescent prominences.
It should be noted that the fast dissipation of large-amplitude prominence oscillations 
is clearly evident in the observational data (e.g., Terradas et al. 2002; Arregui, Oliver and Ballester, 2012, and references cited therein).
In the present case, we specifically found and quantified that the wave-leakage process is the dissipative 
agent for the antisymmetric magnetoacoustic--gravity oscillations in Pikelner's model prominence. 
Since in our model
we do not invoke any non-adiabatic thermodynamical effects, {\it e.g.}, radiative cooling, for the 
dissipation of such oscillations, even then the wave-leakage turns out to be a
very 
effective 
mechanism for the dissipation of asymmetric oscillations. The study of the relative significance of 
various dissipative agents on the magnetoacoustic--gravity mode oscillations, {\it e.g.}, wave-leakage, radiative cooling, {\it etc.} may be 
an important task that we will study in a future project.

In conclusion, we have tested our new analytical prominence model numerically,
and excited the slow magnetoacoustic--gravity oscillations
along its magnetic dip by perturbing a gas pressure within
the prominence. Our model will be further used in a more detailed study
of prominence dynamics and to constrain its oscillations
to exploit prominence seismology in order to deduce local plasma conditions.

\noindent
{\bf Acknowledgements}\hspace{8 mm}This work was supported by the project from the Polish National
Foundation (NCN) Grant no. 2014/15/B/ST9/00106. 
The work has also been supported
by a Marie Curie International Research Staff Exchange
Scheme Fellowship within the 7th European Community
Framework Program (K. Murawski, A. Solov'ev and J. Kra\'skiewicz.). 
In addition, A. Solov'ev thanks the Russian Scientific Foundation for the support in the frame of project No 15-12-20001".
The software used in this work was in part developed
by the DOE-supported ASCI/Alliance Center for
Astrophysical Thermonuclear Flashes at the University of
Chicago. The visualizations of the simulation variables have been
carried out using the IDL (Interactive Data Language) software package.
Numerical simulations were performed on the Solaris cluster 
at Institute of Mathematics of M. Curie-Sk{\l}odowska University in
Lublin, Poland.

\end{article}
\end{document}